\documentclass[preprint,nocomments,norevision]{ptephy_om}
\preprintnumber{RIKEN-iTHEMS-Report-26}
\revisionnum{2}

\usepackage{tikz}
\usepackage{booktabs}

\usepackage[strict]{changepage}
\newcommand{\pagehspace}{%
  \checkoddpage
  \ifoddpage
    \hspace*{-5em}%
  \else
    \hspace*{-4em}%
  \fi
}

\usepackage{comment}

\usepackage{hyperref}
\usepackage{orcidlink}

\title{Quasinormal modes and continuum response of de~Sitter black holes via complex scaling method}

\author{Shoya Ogawa \orcidlink{0000-0003-0900-2486}}
\affil{Department of Physics, Kyushu University, 744 Motooka, Nishi-ku,
Fukuoka 819-0395, Japan
\email{ogawa.shoya.615@m.kyushu-u.ac.jp}
}

\author[2]{Okuto Morikawa \orcidlink{0000-0002-0044-4491}}
\affil[2]{Center for Interdisciplinary Theoretical and Mathematical Sciences (iTHEMS),
RIKEN, Wako 351-0198, Japan
\email{okuto.morikawa@riken.jp}
}

\author[3]{Takuya Hirose \orcidlink{0000-0003-0962-8884}}
\affil[3]{Faculty of Science and Engineering, Kyushu Sangyo University,
Fukuoka 813-8503, Japan
\email{t.hirose@ip.kyusan-u.ac.jp}
}

\begin{document}
\begin{abstract}
We apply the complex scaling method to black-hole perturbations in
four-dimensional Schwarzschild--de~Sitter (dS) spacetimes.
The method converts the outgoing-wave boundary-value problem into a
non-Hermitian spectral problem and enables quasinormal-mode poles and the
rotated continuum to be treated in a common framework.
\revtwo{We ask how the global spectral response is redistributed between
isolated QNM poles and the complementary spectrum as the cosmological constant
is varied.  To address this question, we use the continuum level density (CLD),
a reference-subtracted resolvent trace related to the derivative of a global
scattering phase, and decompose it into selected QNM-pole terms and a
well-defined remainder.}
Using Regge--Wheeler-type perturbation equations for scalar, electromagnetic,
and gravitational fields, we investigate how a nonzero cosmological constant
modifies the pole and continuum sectors.
\revtwo{For the dS$_4$ examples, the lowest QNM provides a natural frequency
and width scale for the associated spectral feature, while the complementary
remainder cannot in general be neglected as $\Lambda$ increases; for $l=3$,
the first overtone contributes visibly in the sharp-feature region.  A
representative numerical-stability test supports the location and overall
shape of the CLD feature under the tested changes of scaling angle and
Gaussian-basis parameters, while its pointwise amplitude remains more
sensitive.  Thus, QNM trajectories and the global spectral response are
related but not interchangeable.}
\revtwo{We also present exploratory extensions to string-inspired
coupled-channel systems and higher-dimensional dS black holes.}
\end{abstract}
\maketitle

\tableofcontents

\section{Introduction}

Black-hole quasinormal modes (QNMs) characterize the linear response of a
black hole to external perturbations and play a central role in black-hole
perturbation theory and gravitational-wave ringdown
physics~\cite{Leaver:1986gd,Ching:1994bd}.
Mathematically, they are resonances and are therefore naturally associated
with the analytic structure of the resolvent or Green function in the complex
frequency plane, rather than with ordinary normalizable
eigenstates~\cite{Leaver:1986gd,Berti:2009kk,Konoplya:2011qq,Casals:2013mpa,Su:2026fvj,Onozawa:1995vu,Richartz:2015saa,Daghigh:2024wcl,Ansorg:2016ztf,Jaramillo:2020tuu,PanossoMacedo:2024nkw,Miyachi:2025ptm,Pombo:2025urp,Motohashi:2024fwt,PanossoMacedo:2025xnf,Takahashi:2025uwo}.
In our recent work~\cite{Ogawa:2026veu}, we developed a complex-scaling-method (CSM) framework~\cite{Aguilar:1971ve,Balslev:1971vb,Simon1972BalslevCombes,
Simon1979ECS,Moiseyev:1998gjp,Myo:2014ypa} for Schwarzschild and
Reissner--Nordstr\"om black holes.
For perturbation equations of Schr\"odinger type in the tortoise coordinate~$r_*$ ($f(r)$ denotes the metric function),
\begin{align}
    \left[
        -\frac{\rmd^2}{\rmd r_*^2}+V(r)
    \right]\psi_l(r_*)
    =
    \omega^2 \psi_l(r_*),
    \qquad
    \frac{\rmd}{\rmd r_*}=f(r)\frac{\rmd}{\rmd r} ,
\end{align}
complex scaling, $r_*\mapsto r_* e^{\rmi\theta}$, turns the outgoing-wave
boundary-value problem into a non-Hermitian spectral problem, in which
resonance poles appear as discrete eigenvalues of the complex-scaled
operator.\footnote{%
Complex-scaling ideas were also used by Vanzo and Zerbini~\cite{Vanzo:2004fy} in the asymptotic analysis of QNMs for multi-horizon black holes.
\revtwo{The present work uses the finite-basis CSM not only to extract
low-lying QNM poles but also to follow the accompanying reference-subtracted
global spectral density.}}
This gives a formulation that stays close to the resonance-theoretic
structure of the problem and is less directly tied to special Frobenius-type
expansions than Leaver-type constructions~\cite{Leaver:1986gd}.

\revtwo{The present work asks a more specific question than the extraction of
QNM frequencies alone: when the cosmological constant changes the QNM spectrum,
how is the associated global spectral response redistributed between the
selected resonance poles and the remainder of the spectrum?}
In four-dimensional Schwarzschild--(anti-)de~Sitter [(A)dS], the metric function becomes
\begin{align}
    f(r)=1-\frac{2M}{r}-\frac{\Lambda}{3}r^2,
    \label{eq:ads_metric_fn}
\end{align}
and the odd-parity Regge--Wheeler-type potential is generalized to
\begin{align}
    \vsup{V}{(odd)}_s(r)
    &=
    f(r)
    \left(
        \frac{l(l+1)}{r^2}
        +\frac{2\sigma M}{r^3}
        -\frac{(s-1)(s-2)\Lambda}{3}
    \right),
    \label{eq:ads_pot_odd}
\end{align}
with $\sigma=1-s^2$ and $n=(l-1)(l+2)/2$.\footnote{%
The even-parity potential for four-dimensional Schwarzschild--(A)dS is also given as
\begin{align}
    \vsup{V}{(even)}_{s=2}(r)
    &=
    \frac{2f(r)}{r^3}
    \frac{
        9M^3+3n^2Mr^2+n^2(n+1)r^3+3M^2(3nr-\Lambda r^3)
    }{
        (3M+nr)^2
    },
    \label{eq:ads_pot_even}
\end{align}
}
Here $s=0$, $1$, and $2$ correspond to scalar, electromagnetic, and
gravitational perturbations, respectively.
Thus, at the level of the master equation, the extension from Schwarzschild is
rather direct.
Schwarzschild--dS black holes provide a standard setting in which the
cosmological constant~$\Lambda$ modifies the spectrum in a qualitatively nontrivial
way~\cite{Zhidenko:2003wq,Choudhury:2003wd,Choudhury:2011at}.\footnote{%
Schwarzschild--AdS and Reissner--Nordstr\"om--AdS black holes have been studied
extensively in the QNM literature~\cite{Berti:2003ud,Horowitz:1999jd,Cardoso:2001bb,Konoplya:2002zu}.}
\revtwo{The four-dimensional Schwarzschild--dS problem is therefore a useful
test bed for separating two pieces of information that are often discussed
together: the motion of isolated QNM poles and the reorganization of the global
continuum spectral density.  Higher-dimensional dS backgrounds and
string-inspired coupled-channel systems provide natural tests of the
portability of the same CSM framework, but in this paper they are treated as
extensions rather than as evidence for the main dS$_4$ CLD conclusions.}

\revtwo{The continuum level density is the quantity we use to formulate this
comparison.  It is the trace of a reference-subtracted resolvent and, in the
standard one-dimensional scattering interpretation, is related to the energy
derivative of a global scattering phase.  The trace therefore provides a
compact scalar measure of how the spectrum is rearranged by the black-hole
potential.  It does not retain the two spatial arguments of the retarded Green
function or source-dependent excitation information, and we do not use it here
as a substitute for a waveform or a greybody factor.}

Continuum contributions are expected to be important for the broader spectral
response and are closely related, in asymptotically flat backgrounds, to
branch-cut contributions and late-time tails~\cite{Leaver:1986gd,Ching:1994bd,Casals:2013mpa,Su:2026fvj}.
The CSM is useful for the present question because the isolated resonance poles
and the rotated discretized continuum appear in the same non-Hermitian
spectral problem~\cite{Moiseyev:1998gjp,Myo:2014ypa}.

\revtwo{Accordingly, the main analysis has three steps.  
First, we identify the low-lying QNM poles for scalar, electromagnetic, and gravitational perturbations and follow their motion with $\Lambda$.  
Second, using the same complex-scaled spectra, we construct the total CLD and separate selected QNM pole terms from the precisely defined remainder.  
This tests whether the location, width scale, and qualitative form of the total spectral feature can be inferred from the selected poles alone.  
Third, because the CLD is a difference of two finite-basis resolvent traces, we test the numerical stability of the CLD directly rather than inferring it from the stability of isolated QNM eigenvalues.  
The general CLD construction and a proof-of-principle pole-resolved illustration were already given in Appendix~B of our previous work~\cite{Ogawa:2026veu}; the present contribution is the systematic Schwarzschild--dS analysis and its dependence on $\Lambda$ across the three perturbation sectors.}

\revtwo{Our numerical conclusions are deliberately matched to the stability
information available in the present data.  We use stable peak positions and
overall spectral shapes to discuss how the pole and remainder sectors are
reorganized, while pointwise amplitudes are treated only qualitatively where a
visible residual basis dependence remains.  The coupled-channel and
higher-dimensional calculations are presented separately as exploratory
extensions of the framework.}

\revtwo{This paper is organized as follows.
In Sec.~\ref{sec:setup}, we summarize the complex-scaling setup and define the
CLD and its numerical interpretation.
In Sec.~\ref{sec:result}, we first present the QNM spectrum and then analyze the
$\Lambda$-dependent CLD, its pole--remainder decomposition, and the numerical
stability of that decomposition.
In Sec.~\ref{sec:discussion}, we state what can and cannot be inferred from the
CLD and relate it to the global scattering phase and greybody factors.
Section~\ref{sec:conclusion} summarizes the conclusions.  
}

\section{Setup}\label{sec:setup}

\subsection{Complex scaling and non-Hermitian spectral problem}
\label{sec:setup-csm}

The complex scaling method is implemented by rotating the tortoise coordinate
as
\begin{align}
    r_* \mapsto r_* e^{\rmi\theta},
\end{align}
with a real scaling angle $\theta$.
Under this transformation, the original outgoing-wave boundary-value problem
is converted into a non-Hermitian spectral problem for the complex-scaled
operator,
\begin{align}
    H^\theta \psi_l^\theta = \omega^2 \psi_l^\theta .
\end{align}
The essential point is that resonance poles, which are defined through
outgoing boundary conditions in the original problem, appear as discrete
eigenvalues of $H^\theta$ after complex scaling.
Note that $\omega$ for any resonance state does not depend on~$\theta$.
This provides a common spectral framework in which pole and continuum sectors
may be discussed simultaneously.

\subsection{Basis expansion and generalized eigenvalue problem}
\label{sec:setup-basis}

To solve the complex-scaled problem numerically, we expand the wave function
in a finite set of $L^2$~basis functions,
\begin{align}
    \psi^\theta(r_*)
    =
    \sum_i c_i \phi_i(r_*).
\end{align}
The explicit choice of basis functions has been considered in the numerical
analysis of Ref.~\cite{Ogawa:2026veu}, and it may be natural to choose the Polynomial $\times$ real range Gaussian basis
\begin{align}
    \phi_{i,n}(r_*)
    \propto
    (\sqrt{\alpha_i}\, r_*)^n e^{-\alpha_i r_*^2},
    \qquad
    \text{$n=0$, $1$, $2$, \dots, $p_{\max}$} .
\end{align}
Here, $\alpha_i = 1/r_i^2$, and
the Gaussian ranges in $[r_{*0},r_{*\max}]$ are taken in geometric progression,
\begin{align}
    r_i = r_{*0}\, a^{\,i-1},
    \qquad
    a = \left( \frac{r_{*\max}}{r_{*0}} \right)^{1/(i_{\max}-1)},
    \label{eq:gaussian-ranges}
\end{align}

With this expansion, the complex-scaled equation is reduced to a matrix
eigenvalue problem of the form
\begin{align}
    \sum_j
    \left(
        H^\theta_{ij}
        -\omega^2 N_{ij}
    \right)c_j
    =0,
\end{align}
where $H^\theta_{ij}$ and $N_{ij}$ denote the Hamiltonian and overlap matrices,
respectively.
The resulting eigenvalues provide the basic spectral data used in the
subsequent QNM and CLD analyses.

\subsection{Continuum level density}
\label{sec:setup-cld}

Besides the isolated resonance poles identified with QNM frequencies, the
complex-scaled spectrum also contains information on the continuum sector.
A useful quantity for this purpose is the continuum level density (CLD), which
measures how the continuum is modified by the interaction relative to a
reference problem.

Let $H$ be the Hamiltonian of the black-hole perturbation problem and let
$H_0$ be a suitable reference (non-interacting) Hamiltonian.
In the present work, the most natural choice is the asymptotic free operator,
so that the CLD describes the deformation of the continuum induced by the
black-hole potential.
The CLD is defined by
\begin{align}
    \Delta \rho(E)
    =
    -\frac{1}{\pi}
    \im
    \Tr
    \left[
        \frac{1}{E-H+\rmi 0}
        -
        \frac{1}{E-H_0+\rmi 0}
    \right],
    \label{eq:cld_definition}
\end{align}
where $E$ denotes the spectral parameter \cite{Suzuki:2005wv,Avishai:1985,Myo:2014ypa}.
In the present Schr\"odinger-type problem, we identify
\begin{align}
    E=\omega^2 .
\end{align}

After complex scaling, one may equivalently evaluate the CLD from the
complex-scaled operators,
\begin{align}
    \Delta \rho^\theta(E)
    =
    -\frac{1}{\pi}
    \im
    \Tr
    \left[
        \frac{1}{E-H^\theta}
        -
        \frac{1}{E-H_0^\theta}
    \right],
    \label{eq:cld_csm}
\end{align}
where $H^\theta$ and $H_0^\theta$ denote the complex-scaled Hamiltonians.
In an exact treatment, the resulting CLD is independent of the scaling angle
within the admissible range, while in a finite-basis calculation a residual
$\theta$ dependence remains and may be used as a practical measure of
numerical stability.
\revtwo{For a finite-basis calculation, the stability of an isolated QNM
eigenvalue does not by itself establish the stability of the CLD.  The latter
involves a cancellation between the full and reference resolvent traces, and
can therefore be more sensitive to the scaling angle, basis truncation, and
radial range.  In Sec.~\ref{sec:cld-validation} we test the
reference-subtracted trace itself under the parameter variations actually used
in the calculation, with identical finite trial spaces and numerical settings
for $H^\theta$ and $H_0^\theta$ in each comparison.}

\revtwo{The trace in Eq.~\eqref{eq:cld_definition} is a global density of states.
For the purpose of this paper, the physically relevant question is whether its
spectral features can be organized in terms of the QNM poles that are visible
in the same complex-scaled spectrum, and how the complementary remainder
changes as $\Lambda$ is varied.  We therefore distinguish robust statements
about the location and qualitative shape of a feature from stronger
pointwise statements about its amplitude.  In the standard scattering
interpretation the CLD is related to the derivative of a global scattering
phase, as reviewed in Sec.~\ref{sec:discussion-greybody}; it is not a
spatially resolved Green function or a waveform observable.}

In asymptotically flat black holes, the continuum sector is closely related to
branch-cut contributions and late-time tails.  Here we use the CLD more
modestly as a reference-subtracted global spectral diagnostic for the dS$_4$
problem within the same CSM calculation that yields the QNM poles.

For this reason, we leave the construction of an appropriate reference
Hamiltonian $H_0$ for the AdS case, compatible with the AdS boundary
condition, as an interesting and highly subtle problem for future work.

\section{QNM and CLD for dS black holes}\label{sec:result}

\subsection{Numerical results of QNMs}

We first present the QNM frequencies obtained by applying the present CSM
framework to four-dimensional Schwarzschild--dS black holes.
The purpose of this subsection is twofold.
First, we check that the complex-scaled formulation can identify isolated
resonance eigenvalues also in the presence of a nonzero cosmological constant.
Second, we examine how the low-lying QNM spectrum changes as the cosmological
constant $\Lambda$ is varied.

We consider scalar, electromagnetic, and gravitational perturbations,
corresponding to $s=0$, $s=1$, and $s=2$, respectively.
For each spin sector, we compute two representative angular momenta,
which we denote by $l=2$ and $l=3$.
The notation $l$ is used throughout this paper for the angular momentum
quantum number.
For each value of $\Lambda$, the corresponding Schwarzschild--dS
Regge--Wheeler-type potential is inserted into the complex-scaled
Schr\"odinger equation, and the QNM candidates are extracted as isolated
eigenvalues separated from the rotated continuum.
The identification of QNM candidates is performed by looking for isolated
eigenvalues that remain stable under changes of the numerical parameters.

In the numerical analysis below, we focus on the low-lying modes.
As in the Schwarzschild and Reissner--Nordstr\"om analysis of our previous
work, these modes are expected to be the most stable against changes of the
basis size, basis parameters, integration range, and scaling angle.
By contrast, higher overtones are typically broader and lie closer to the
rotated continuum, making their identification more sensitive to numerical
details.
For this reason, the present calculation should be understood as a first
systematic test of the CSM framework for Schwarzschild--dS backgrounds,
rather than as a final optimized computation of all overtones.

For each channel, we perform calculations for several values of the
cosmological constant.
Positive values,
\begin{align}
    \Lambda>0,
\end{align}
correspond to Schwarzschild--dS backgrounds.
The limit $\Lambda\to0$ connects the present setup to the asymptotically flat
Schwarzschild problem and therefore serves as an important reference point.
Particular attention will be paid to how the real and imaginary parts of the
QNM frequencies move as $\Lambda$ is varied.

\begin{figure}[htbp]
 \centering
 \begin{tikzpicture}
  \node at (0,8) {\includegraphics[width=0.45\linewidth]{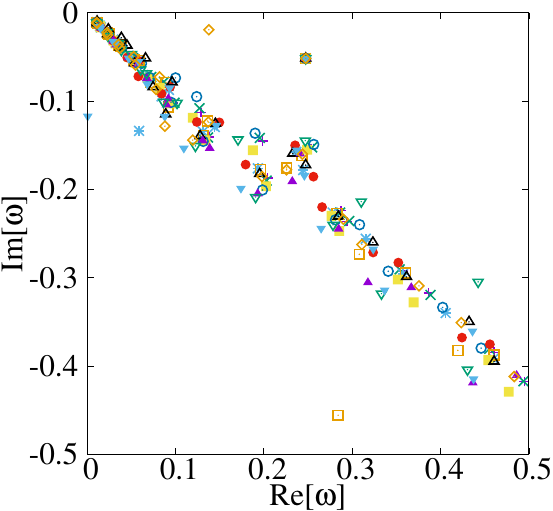}};
  \node at (8,8) {\includegraphics[width=0.45\linewidth]{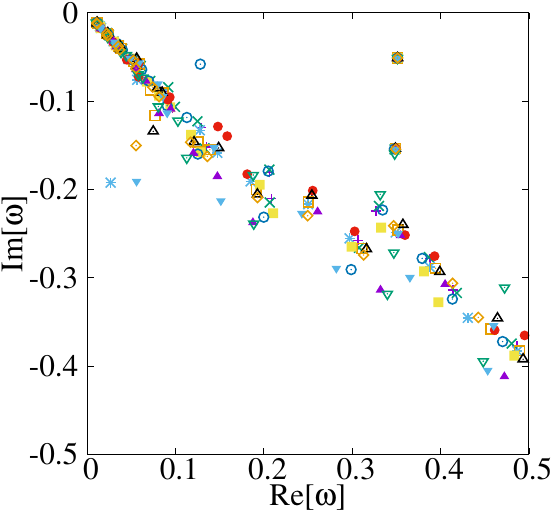}};
  \node[draw, rectangle] at (-1,8) {$l=2$};
  \node[draw, rectangle] at (7,8) {$l=3$};
  \draw[thick, red] (0.425,10.65) circle (0.3);
  \node[above] at (1.2,10.6) {$n=0$};
  \draw[thick, red] (9.65,10.65) circle (0.3);
  \node[above] at (10.655,10.7) {$n=0$};
  \draw[thick, red] (9.55,9.45) circle (0.3);
  \node[above] at (10.45,9.2) {$n=1$};
 \end{tikzpicture}
 \caption{The Schwarzschild--dS case with $s=0$. $M=1.0$.
 The discretized continuum spectrum is represented by the sequence of points aligned approximately along the $\theta$-rotated direction.
 The isolated points on the real-axis side are the QNM frequencies.
 Eigenvalues of the complex-scaled Hamiltonian $H^\theta$ for $l=2$ and $l=3$ are shown.
 The scaling angle is taken to be~$\theta=42^\circ$--$44^\circ$.
 In these figures, we fix the cosmological-constant parameter to $\Lambda=0.08$.
 The eigenvalues enclosed by red circles are identified as QNM candidates, since they appear as isolated points separated from the rotated continuum and remain stable under moderate changes of the scaling angle and basis parameters.}
 \label{fig:qnm_ds_s0}
\end{figure}

\begin{figure}[htbp]
 \centering
 \begin{tikzpicture}
  \node at (0,8) {\includegraphics[width=0.45\linewidth]{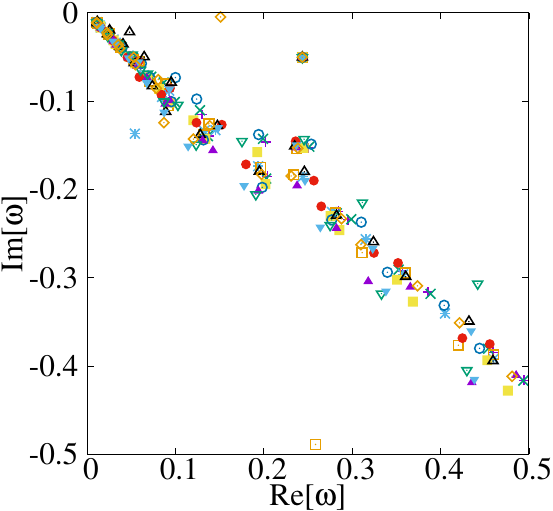}};
  \node at (8,8) {\includegraphics[width=0.45\linewidth]{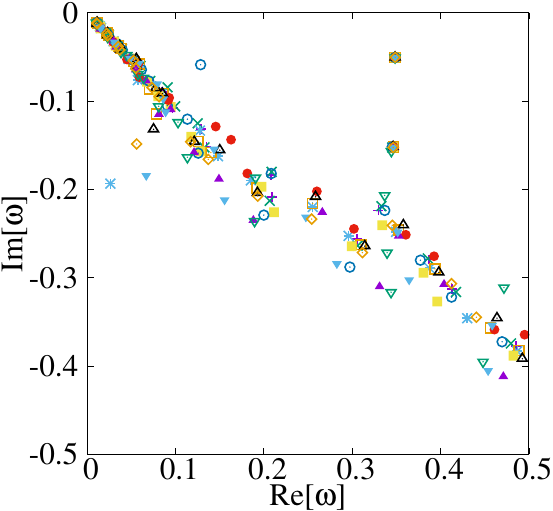}};
  \node[draw, rectangle] at (-1,8) {$l=2$};
  \node[draw, rectangle] at (7,8) {$l=3$};
  \draw[thick, red] (0.4,10.7) circle (0.3);
  \node[above] at (1.625,10.6) {$n=0$};
  \draw[thick, red] (9.6,10.65) circle (0.3);
  \node[above] at (10.655,10.7) {$n=0$};
  \draw[thick, red] (9.55,9.45) circle (0.3);
  \node[above] at (10.45,9.2) {$n=1$};
 \end{tikzpicture}
 \caption{The Schwarzschild--dS case with $s=1$. $M=1.0$.}
 \label{fig:qnm_ds_s1}
\end{figure}

\begin{figure}[htbp]
 \centering
 \begin{tikzpicture}
  \node at (0,8) {\includegraphics[width=0.45\linewidth]{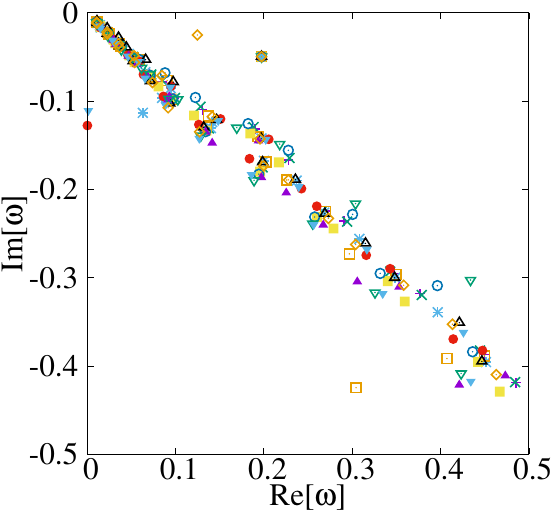}};
  \node at (8,8) {\includegraphics[width=0.45\linewidth]{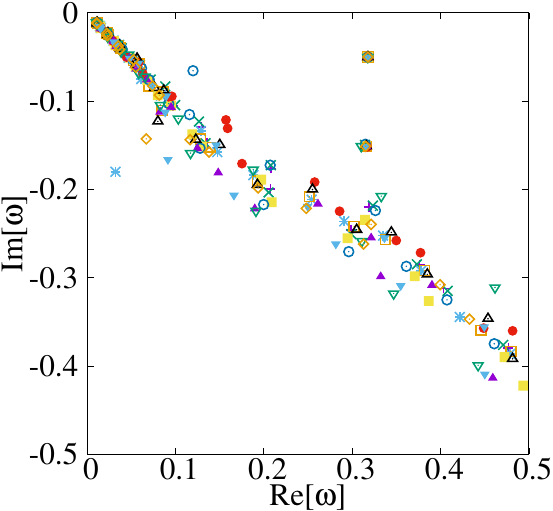}};
  \node[draw, rectangle] at (-1,8) {$l=2$};
  \node[draw, rectangle] at (7,8) {$l=3$};
  \draw[thick, red] (-0.2,10.65) circle (0.3);
  \node[above] at (0.8,10.6) {$n=0$};
  \draw[thick, red] (9.25,10.65) circle (0.3);
  \node[above] at (10.2,10.7) {$n=0$};
  \draw[thick, red] (9.2,9.5) circle (0.3);
  \node[above] at (10.1,9.2) {$n=1$};
 \end{tikzpicture}
 \caption{The Schwarzschild--dS case with $s=2$. $M=1.0$.}
 \label{fig:qnm_ds_s2}
\end{figure}

The numerical results are summarized in
Figs.~\ref{fig:qnm_ds_s0}--\ref{fig:qnm_ds_s2}
for Schwarzschild--dS.
Here
Fig.~\ref{fig:qnm_ds_s0} shows the scalar results,
Fig.~\ref{fig:qnm_ds_s1} shows the electromagnetic results, and
Fig.~\ref{fig:qnm_ds_s2} shows the gravitational results.
In these figures, we fix the magnitude of the cosmological constant to
$\Lambda=0.08$.
For each channel, we combine the results obtained for several values of the
scaling angle,
\begin{align}
    \theta = 42^\circ\text{--}44^\circ ,
    \label{eq:search_theta}
\end{align}
and for the Gaussian parameter sets
\begin{align}
    i_{\max}=30,\ 40,
    \qquad
    r_{*0}=0.1,
    \qquad
    r_{*\max}=60,\ 80 .
    \label{eq:search_gaussian_param}
\end{align}
The complex eigenvalues of the complex-scaled Hamiltonian are displayed in the
complex $\omega$ plane.
The QNM candidates are identified as isolated eigenvalues that remain stable
under these moderate changes of the scaling angle and basis parameters.

The QNM candidates are identified as the eigenvalues enclosed by the red
circles in Figs.~\ref{fig:qnm_ds_s0}--\ref{fig:qnm_ds_s2}. These eigenvalues
appear as isolated points separated from the rotated continuum and remain
stable under moderate changes of the scaling angle and basis parameters.
We find that the resulting frequencies agree with the previous results of
Ref.~\cite{Zhidenko:2003wq} with high accuracy, which provides a nontrivial validation of
the present CSM identification procedure.

Table~\ref{tab:result_ds} presents the numerical
results obtained by the CSM for the Schwarzschild--dS case.
In this table, we list the lowest QNM frequencies, $n=0$, for scalar
($s=0$), electromagnetic ($s=1$), and gravitational ($s=2$) perturbations.
For each spin sector, the results for $l=2$ and $l=3$ are shown for several
values of the cosmological constant~$\Lambda$.
The numbers in parentheses indicate the numerical uncertainty estimated from
the spread of the results obtained with different values of the scaling angle
and basis parameters.\footnote{%
To estimate the numerical uncertainty of the extracted QNM frequencies, we
use the spread of the results obtained from different numerical parameter
sets.
Let $\omega_i$ denote the frequency obtained in the $i$th run, and let
$\vsub{N}{samp}$ be the number of runs.
We define the averaged frequency by
\begin{align}
    \bar{\omega}
    =
    \frac{1}{\vsub{N}{samp}}
    \sum_{i=1}^{\vsub{N}{samp}}
    \omega_i ,
\end{align}
and estimate the numerical uncertainty as
$\Delta\omega=\max_i\left|\omega_i-\bar{\omega}\right|$.
We then quote the result as
$\omega=\Bar{\omega}\pm\Delta\omega$.
The quantity $\Delta\omega$ should be understood as a practical measure of
numerical stability, not as a statistical error bar.}

\begin{table}[htbp]
\centering
\caption{Numerical results of lowest QNM frequencies ($n=0$) of Schwarzschild--dS black hole for $l=2$ and $l=3$.
The entries are given as
$\omega=\re\omega-\rmi|\im\omega|$.
For comparison, the table also includes the Schwarzschild results obtained
from a separate calculation ($\Lambda=0$).
The numbers in parentheses indicate the numerical uncertainty estimated from
the spread under changes of the scaling angle and basis parameters.}
\label{tab:result_ds}
\begin{tabular}{l
                l@{$-\rmi$}l
                l@{$-\rmi$}l}
\toprule
$\Lambda$ & \multicolumn{2}{c}{$l=2$} & \multicolumn{2}{c}{$l=3$} \\
\midrule 
\multicolumn{5}{c}{$s=0$} \\
$0$    & $0.483646(91)$     & $0.09676(15)$
       & $0.67539(23)$ & $0.09654(11)$ \\
$0.02$ & $0.43462(38)$ & $0.08859(15)$
       & $0.60907(12)$ & $0.08789(25)$ \\
$0.04$ & $0.38078(15)$ & $0.07875(10)$
       & $0.53583(12)$ & $0.07789(16)$ \\
$0.06$ & $0.320020(55)$ & $0.066838(67)$
       & $0.45235(10)$ & $0.06608(33)$ \\
$0.08$ & $0.247470(18)$ & $0.051904(44)$
       & $0.351393(55)$ & $0.051402(57)$ \\
$0.10$ & $0.146608(13)$ & $0.030685(18)$
       & $0.209096(16)$ & $0.030556(24)$ \\
\addlinespace
\multicolumn{5}{c}{$s=1$} \\
$0$    & $0.457595(89)$ & $0.09500(11)$
       & $0.65692(27)$ & $0.095648(80)$ \\
$0.02$ & $0.41503(28)$ & $0.08614(14)$
       & $0.59527(13)$ & $0.08665(23)$ \\
$0.04$ & $0.36723(12)$ & $0.076233(92)$
       & $0.52630(12)$ & $0.07662(15)$ \\
$0.06$ & $0.311815(48)$ & $0.064772(63)$
       & $0.44657(17)$ & $0.06505(30)$ \\
$0.08$ & $0.243642(13)$ & $0.050673(45)$
       & $0.348668(52)$ & $0.050791(56)$ \\
$0.10$ & $0.145816(11)$ & $0.030372(18)$
       & $0.208523(17)$ & $0.030400(23)$ \\
\addlinespace
\multicolumn{5}{c}{$s=2$} \\
$0$    & $0.37366(24)$ & $0.08896(15)$
       & $0.59947(23)$ & $0.09270(28)$ \\
$0.02$ & $0.338393(88)$ & $0.081754(78)$
       & $0.54308(14)$ & $0.08451(19)$ \\
$0.04$ & $0.298896(51)$ & $0.073288(68)$
       & $0.480033(87)$ & $0.07517(26)$ \\
$0.06$ & $0.253287(44)$ & $0.063047(71)$
       & $0.40718(19)$ & $0.064137(89)$ \\
$0.08$ & $0.197474(19)$ & $0.049874(42)$
       & $0.317804(32)$ & $0.050378(49)$ \\
$0.10$ & $0.117916(21)$ & $0.030209(21)$
       & $0.189991(12)$ & $0.030313(21)$ \\
\bottomrule
\end{tabular}
\end{table}

For the Schwarzschild--dS case shown in Table~\ref{tab:result_ds},
both \(\re\omega\) and \(|\im\omega|\) decrease as the positive cosmological
constant \(\Lambda\) is increased.
This tendency is observed in all spin sectors considered here and for both
angular momenta \(l=2\) and \(l=3\).
It is consistent with the approach to the near-Nariai regime, in which the
black-hole and cosmological horizons become closer and the characteristic
oscillation and damping scales are reduced.

We also list the Schwarzschild results obtained from a separate calculation
for comparison.
For small \(\Lambda\), the low-lying Schwarzschild--dS QNM frequencies vary
smoothly from the Schwarzschild values.
The QNM candidates selected from the complex-scaled spectra agree with the
previous results of Ref.~\cite{Zhidenko:2003wq} with high accuracy.
This agreement provides a nontrivial validation of the present CSM
identification procedure: the stable isolated eigenvalues separated from the
rotated continuum are correctly identified with the physical QNM poles.

The quoted uncertainties are estimated from the spread of the results under
moderate changes of the scaling angle and basis parameters.
They should therefore be understood as practical measures of numerical
stability, rather than as statistical error bars.
The small uncertainties of the low-lying modes indicate that these resonance
eigenvalues are robustly extracted within the present finite-basis CSM
calculation.

These QNM results serve as a benchmark for the subsequent CLD analysis.
While the QNM frequencies characterize the isolated pole sector, the CLD
probes how the continuum part of the spectral response is modified by the
black-hole potential.
In the following subsection, we therefore use the same complex-scaled
framework to analyze the continuum level density for the Schwarzschild--dS
background.

\subsection{\revtwo{Continuum level density and pole--remainder reorganization}}

We next analyze the continuum level density for the Schwarzschild--dS
backgrounds using the same complex-scaled spectra as those used for the QNM
analysis.
The CLD introduced in Eq.~\eqref{eq:cld_csm} is defined as a density with
respect to the spectral variable $E=\omega^2$.
In the following plots, however, we display the corresponding density with
respect to the frequency variable $\omega$.
We therefore define
\begin{align}
    \Delta^\theta(\omega)
    &=
    \frac{\rmd E}{\rmd \omega}
    \Delta\rho^\theta(E)
    \bigg|_{E=\omega^2}
    \notag\\
    &=
    2\omega\,
    \Delta\rho^\theta(\omega^2).
    \label{eq:cld_omega_density}
\end{align}
This Jacobian factor should be kept in mind when comparing the CLD plotted
as a function of~$\omega$ with the original definition in terms of~$E$.

\begin{figure}[ht]
    \pagehspace
    \includegraphics[width=1.25\columnwidth]{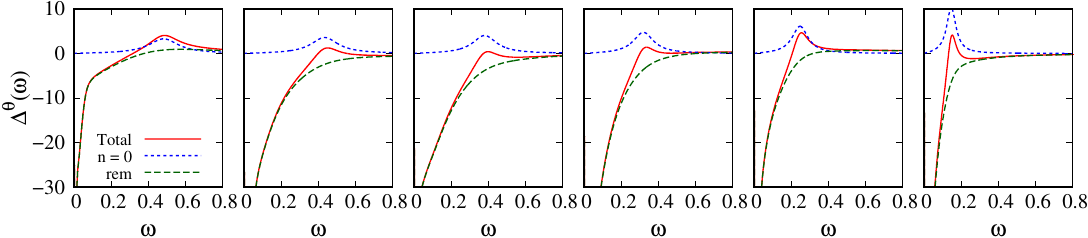}
    \caption{\revone{CLD and its pole--remainder decomposition for $s=0$,
    $l=2$, and $\theta=43^\circ$.  The six panels correspond, from left to
    right, to~$\Lambda=0$, $0.02$, $0.04$, $0.06$, $0.08$, and $0.10$.
    Shown separately are the total CLD, the pure $n=0$ pole contribution of
    Eq.~\eqref{eq:cld_pole_omega}, and the 
    \revtwo{complementary remainder} of Eq.~\eqref{eq:cld_remainder_omega}.}}
    \label{fig:cld_s0_l2}
\end{figure}

\begin{figure}[ht]
    \pagehspace
    \includegraphics[width=1.25\columnwidth]{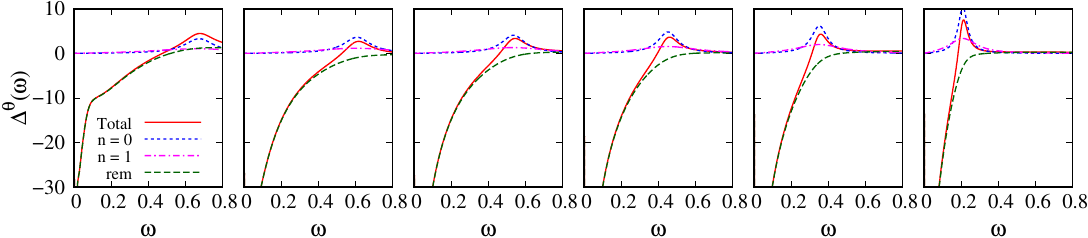}
    \caption{\revone{Same as Fig.~\ref{fig:cld_s0_l2}, but for $s=0$ and $l=3$. 
    \revtwo{The pure $n=1$ pole contribution is also shown separately; both the $n=0$ and $n=1$ contributions are excluded from the remainder.}
    }}
    \label{fig:cld_s0_l3}
\end{figure}

\begin{figure}[ht]
    \pagehspace
    \includegraphics[width=1.25\columnwidth]{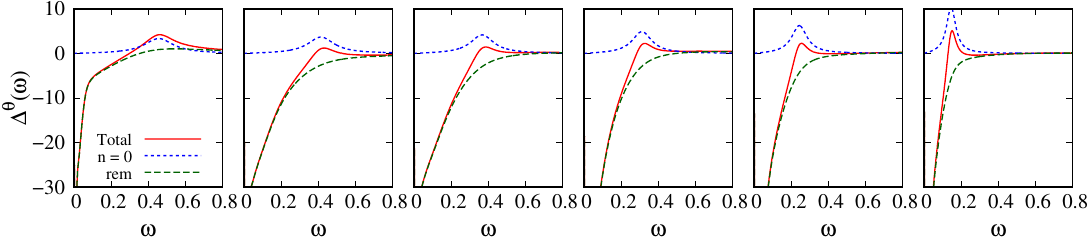}
    \caption{\revone{Same as Fig.~\ref{fig:cld_s0_l2}, but for $s=1$ and
    $l=2$.}}
    \label{fig:cld_s1_l2}
\end{figure}

\begin{figure}[ht]
    \pagehspace
    \includegraphics[width=1.25\columnwidth]{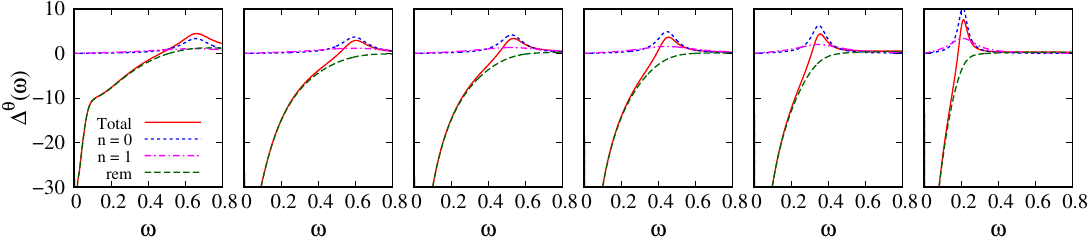}
    \caption{\revone{Same as Fig.~\ref{fig:cld_s0_l3}, but for $s=1$ and
    $l=3$.}}
    \label{fig:cld_s1_l3}
\end{figure}

\begin{figure}[ht]
    \pagehspace
    \includegraphics[width=1.25\columnwidth]{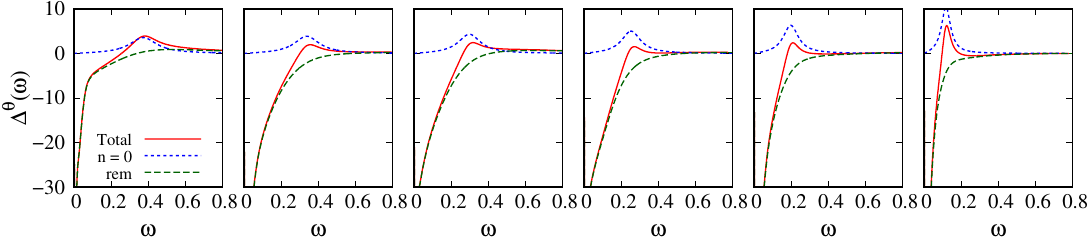}
    \caption{\revone{Same as Fig.~\ref{fig:cld_s0_l2}, but for $s=2$ and
    $l=2$.}}
    \label{fig:cld_s2_l2}
\end{figure}

\begin{figure}[ht]
    \pagehspace
    \includegraphics[width=1.25\columnwidth]{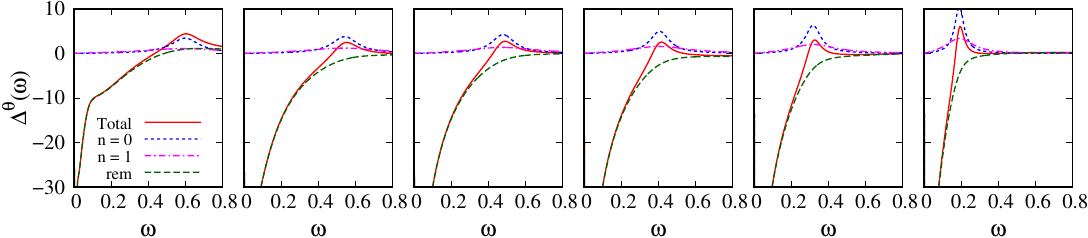}
    \caption{\revone{Same as Fig.~\ref{fig:cld_s0_l3}, but for $s=2$ and
    $l=3$.}}
    \label{fig:cld_s2_l3}
\end{figure}

\revtwo{Figures~\ref{fig:cld_s0_l2}--\ref{fig:cld_s2_l3} show
$\Delta^\theta(\omega)$ together with the selected-pole contributions and the
remainder for scalar, electromagnetic, and gravitational perturbations with
$l=2$ and $l=3$.  The purpose of the comparison is to determine which aspects
of the total spectral feature track an identifiable QNM pole and which require
the rest of the spectrum.}
\begin{align}
    \Lambda
    =
    0,\ 0.02,\ 0.04,\ 0.06,\ 0.08,\ 0.10 .
\end{align}
\revtwo{In each panel we distinguish the total CLD, the pure contributions of the explicitly selected QNM poles, and the complementary remainder. Let $\mathcal{S}_l$ denote the set of overtone indices of the explicitly selected QNM poles. In the present analysis, we take $\mathcal{S}_2=\{0\}$ and $\mathcal{S}_3=\{0,1\}$.}

Let $E^\theta_\nu$ be the complex-scaled eigenvalues of the full Hamiltonian
and let $E^\theta_{0,\nu'}$ be those of the reference Hamiltonian.
The total quantity plotted in the figures is
\begin{align}
    \vsub{\Delta}{tot}^{\theta}(\omega)
    =
    -\frac{2\omega}{\pi}
    \im
    \left[
    \sum_\nu
    \frac{1}{\omega^2-E^\theta_\nu}
    -
    \sum_{\nu'}
    \frac{1}{\omega^2-E^\theta_{0,\nu'}}
    \right].
    \label{eq:cld_total_omega}
\end{align}
\revtwo{Let $E^\theta_n$ denote an eigenvalue associated with an explicitly selected QNM pole. Its pure contribution, with no reference-spectrum term included, is defined as}
\begin{align}
    \revone{\Delta^\theta_{n}(\omega)}
    &\revone{=
    -\frac{2\omega}{\pi}
    \im\left[
    \frac{1}{\omega^2-E^\theta_{n}}
    \right].}
    \label{eq:cld_pole_omega}
\end{align}
\revtwo{We then define the remainder by the exact identity
\begin{align}
    \vsub{\Delta}{rem}^{\theta}(\omega)
    &=
    \vsub{\Delta}{tot}^{\theta}(\omega)
    -
    \sum_{n\in\mathcal{S}_l}
    \Delta^\theta_n(\omega).
    \label{eq:cld_remainder_omega}
\end{align}
The remainder contains all full-Hamiltonian spectral contributions
other than the selected QNM-pole terms, together with the complete
reference subtraction. It is therefore not a purely nonresonant background;
in particular, it may still contain additional QNM poles that are not
separately identified.}

\revtwo{Appendix~\ref{sec:cld-alternative-diagnostic} discusses an alternative reference-subtracted single-pole diagnostic, which coincides with the diagnostic used in Appendix B of Ref.~\cite{Ogawa:2026veu}. 
It answers a different question: how well a single-pole approximation to the interacting resolvent trace reproduces the total reference-subtracted CLD.}

\revtwo{The first robust trend is the migration of the CLD structure with
$\Lambda$.  In the representative scalar $s=0$, $l=2$ sector, the total CLD is
broad at small $\Lambda$ and develops a narrower feature at lower frequency as
$\Lambda$ increases.  The lowest QNM pole moves in the same direction:
Table~\ref{tab:result_ds} shows that both $\re\omega_{n=0}$ and
$|\im\omega_{n=0}|$ decrease toward the near-Nariai regime.  The isolated pole
therefore supplies a natural location and width scale for the evolving
feature.}

\revtwo{The total CLD, however, is not determined by that pole alone.  At small
$\Lambda$, the broad total feature and the pure $n=0$ contribution have similar
locations, signs, and widths, and the remainder is comparatively modest around
the maximum.  For the sharper feature at larger $\Lambda$, the remainder can
no longer be treated as a small correction and has the opposite sign over the
relevant frequency range.  The displayed decomposition therefore indicates a
qualitatively important pole--remainder cancellation.  Because
Sec.~\ref{sec:cld-validation} finds a visible residual dependence in the
pointwise CLD amplitude, we use this comparison only to establish the
qualitative importance and sign of the remainder; we do not infer a precision
cancellation ratio or a converged peak height from the present data.}


\revtwo{The $l=3$ sectors sharpen this conclusion. 
As $\Lambda$ increases, the pure $n=1$ contribution develops a pronounced positive feature in the same frequency region as the sharp structure of the total CLD. 
Thus, the first overtone is not generically a negligible correction in this regime. 
In these panels, both the $n=0$ and $n=1$ pole contributions are separated explicitly from the remainder. 
The remaining contribution is nevertheless non-negligible, showing that the total CLD is not exhausted by the two displayed QNM poles.}

\revtwo{Taken together, the figures answer the main modal question at the level
supported by the present numerics: QNM frequencies provide useful resonance
scales for the locations and widths of spectral features, but the total global
spectral density can require higher poles and the reference-subtracted
continuum sector.  The relative importance of these pieces changes with
$\Lambda$.}

\subsection{\revtwo{Numerical stability and scope of the CLD analysis}}
\label{sec:cld-validation}

\revtwo{The CLD stability test is logically separate from the stability
tests used to identify isolated QNM poles.  As a representative case, we use
the $s=2$, $l=3$, $\Lambda=0.08$ CLD and repeat the calculation while varying
the scaling angle and Gaussian-basis parameters.  The full and reference
Hamiltonians are represented in identical finite trial spaces for every
parameter set.}

\revone{Specifically, we use
\begin{align}
    \theta &= 42^\circ,\ 43^\circ,\ 44^\circ,\\
    i_{\max} &= 30,\ 40,\qquad
    r_{*\max}=60,\ 80,
\end{align}
with the remaining numerical parameters fixed to those used in the
production calculation.  These choices give twelve parameter combinations.}

\begin{figure}[ht]
    \centering
    \includegraphics[width=0.75\linewidth]{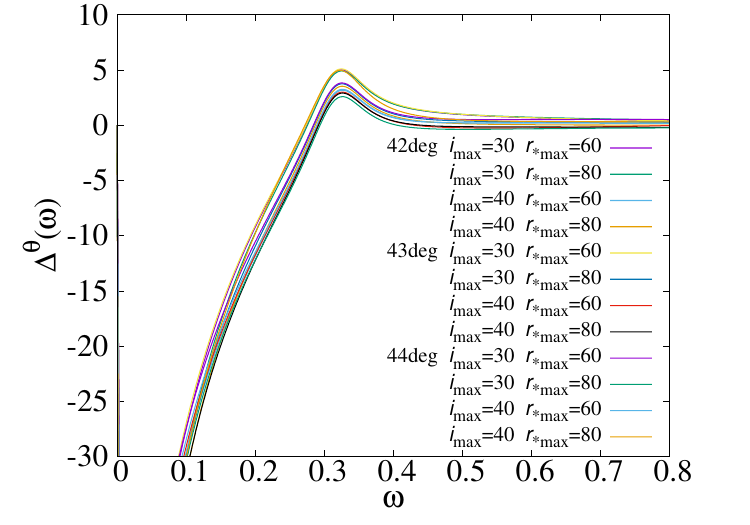}
    \caption{\revone{Numerical stability of the CLD for the representative
    $s=2$, $l=3$ case.  The twelve curves correspond to
    $\theta=42^\circ$, $43^\circ$, and $44^\circ$,
    $i_{\max}=30$ and $40$, and $r_{*\max}=60$ and $80$, $\Lambda=0.08$.
    Identical numerical parameters are used for the full and reference
    Hamiltonians in each calculation.}}
    \label{fig:varying_params}
\end{figure}

\revtwo{Figure~\ref{fig:varying_params} shows a clear hierarchy in the
numerical stability.  The frequency at which the main positive feature occurs
and its broad overall shape are stable under the tested variations.  The
pointwise amplitude is less stable and retains a visible dependence on the
finite-basis parameters.  This is consistent with the fact that the CLD is a
cancellation between two discretized resolvent traces and is more sensitive
than the position of an isolated QNM eigenvalue.}

\revtwo{We therefore use this test to define the scope of the conclusions in
Sec.~\ref{sec:result}: statements based on the position, sign pattern, and
qualitative pole--remainder structure are regarded as numerically supported,
whereas numerical peak heights are not treated as precision observables.  The
present test is restricted to one representative CLD and hence does not
establish uniform pointwise convergence over all broad and sharp structures.
An independent comparison with a real-frequency scattering calculation would
provide a stronger quantitative validation of the finite-basis CLD by testing
whether the integrated CLD reproduces the physical scattering phase.  Such a
benchmark is not part of the present numerical dataset and is left for future
work.}

\section{Discussion}\label{sec:discussion}

\subsection{\revtwo{What is learned from the CLD?}}

\revtwo{The QNM spectrum and the CLD answer different but related questions.
Table~\ref{tab:result_ds} locates isolated resonance poles and shows how their
frequency and damping scales move with $\Lambda$.  The CLD asks how the global
reference-subtracted spectral density is distributed around those scales.  By
constructing both quantities from the same complex-scaled spectrum, the
present calculation makes it possible to ask directly whether a visible
spectral feature is accounted for by one selected QNM or whether the rest of
the spectrum is essential.}

\revtwo{The answer depends on the regime.  For the broad Schwarzschild-like
features, the lowest pole tracks the location and width of the total CLD
reasonably well.  As $\Lambda$ is increased and the feature becomes sharper,
the 
complementary remainder
can no longer be treated as a negligible
correction and contributes with the opposite sign in the displayed
decomposition. 
In the $l=3$ sectors, the explicitly separated first-overtone contribution also becomes appreciable.
This is information that is not contained
in a table of QNM frequencies: the pole locations alone do not tell us how the
total spectral-phase density is partitioned among the selected resonances and
the complementary spectrum.}

\revtwo{The scope of that statement is important.  The CLD is a trace quantity,
so it deliberately discards the spatial and source information retained by a
Green function.  The present result is therefore a statement about global
spectral organization, not about a time-domain waveform or a source-dependent
excitation amplitude.  Likewise, because the representative stability test
shows residual pointwise amplitude dependence, the numerical conclusion is
structural rather than a precision determination of peak heights.}

\revtwo{For the present single-channel radial equation, direct real-frequency
integration remains the simpler route if the only desired output is the total
scattering phase or transmission probability.  The role of the CSM here is
different: the same diagonalization that identifies the QNM poles also gives a
spectral representation in which selected poles and the remainder can be
followed as $\Lambda$ is varied.  This pole-resolved organization is the
specific use of the CLD in the present work; it is not claimed as a universal
accuracy or cost advantage over direct scattering methods.}

\subsection{CLD, transmission phase, and greybody factors}
\label{sec:discussion-greybody}

It is useful to clarify the relation between the present CLD analysis and
greybody factors.
Greybody factors are transmission probabilities associated with black-hole
scattering problems.
They describe how the effective potential outside the black hole modifies the
spectrum of radiation from a perfect blackbody spectrum and are therefore
more directly connected to observable fluxes.
They have been studied in a variety of black-hole backgrounds, including
higher-dimensional, charged, dS, and AdS cases~\cite{Harmark:2007jy}.
More recently, greybody factors have also been discussed in connection with
black-hole ringdown amplitudes and as quantities complementary to the usual
QNM description~\cite{Oshita:2023cjz,Konoplya:2024lir}.

The CLD is not itself a transmission coefficient.
Rather, it measures the change of the continuum density of states relative to
a reference problem.
In the CSM, the CLD~\eqref{eq:cld_csm} is evaluated as
\begin{align}
    \Delta \rho^\theta(E)
    &=
    -\frac{1}{\pi}
    \im
    \Tr
    \left[
        \frac{1}{E-H^\theta}
        -
        \frac{1}{E-H_0^\theta}
    \right]
    \notag\\
    &=
    -\frac{1}{\pi}
    \im
    \left[
        \sum_\nu \frac{1}{E-E^\theta_\nu}
        -
        \sum_{\nu'} \frac{1}{E-E^\theta_{0,\nu'}}
    \right],
    \label{eq:cld_csm_nu}
\end{align}
where $H^\theta$ and $H_0^\theta$ are the complex-scaled full and reference
Hamiltonians, respectively, and
$E^\theta_\nu$ and $E^\theta_{0,\nu'}$ denote their complex eigenvalues.
Equation~\eqref{eq:cld_csm_nu} naturally decomposes the CLD into individual
spectral contributions,
\begin{align}
    \Delta\rho^\theta(E)
    =
    \sum_\nu \rho^\theta_\nu(E)
    -
    \sum_{\nu'} \rho^\theta_{0,\nu'}(E).
\end{align}
\revtwo{Equation~\eqref{eq:cld_csm_nu} makes the pole--remainder
organization used above transparent: isolated eigenvalues can be displayed as
individual resonance-pole terms, while all remaining full-spectrum terms and
the reference subtraction form the complementary contribution.  This is an
exact regrouping of the finite-basis trace once the selected poles are fixed;
the physical content comes from comparing how these pieces reorganize the same
total CLD as $\Lambda$ is varied.}

In ordinary one-dimensional scattering theory, the change in the density of
states is related to the energy derivative of the phase of the transmission
amplitude~\cite{Avishai:1985}.
Writing
\begin{align}
    t(E)=|t(E)|e^{i\delta(E)},
\end{align}
one obtains, up to the conventional normalization of the phase,
\begin{align}
    \Delta\rho(E)
    =
    \frac{1}{\pi}
    \frac{\rmd \delta(E)}{\rmd E}.
    \label{eq:cld_phase_relation}
\end{align}
Equivalently,
\begin{align}
    \delta(E)
    =
    \pi
    \int^E \rmd E'\,\Delta\rho(E')
    +
    \delta_0 ,
    \label{eq:phase_from_cld}
\end{align}
where $\delta_0$ is an energy-independent constant fixed by the choice of
reference phase.
Using the CSM decomposition of Eq.~\eqref{eq:cld_csm_nu}, this phase may be
written as
\begin{align}
    \delta(E)
    =
    \sum_\nu \delta_\nu(E)
    -
    \sum_{\nu'} \delta_{0,\nu'}(E)
    +
    \delta_0 ,
\end{align}
with
\begin{align}
    \delta_\nu(E)
    =
    \pi\int^E \rmd E'\,\rho^\theta_\nu(E').
\end{align}
\revtwo{Equations~\eqref{eq:cld_phase_relation} and
\eqref{eq:phase_from_cld} provide the scattering interpretation used in this
paper: the CLD is a density for the global scattering phase, and the spectral
decomposition of the CLD induces a corresponding decomposition of that phase.
This relation identifies what the traced quantity measures.  The present
calculation does not independently reconstruct the real-frequency phase, so we
do not use these equations as a numerical validation of the finite-basis CLD.}

This observation suggests a possible bridge to greybody factors.
For a black-hole scattering problem, the greybody factor is essentially a
transmission probability,
\begin{align}
    \Gamma(E)=|t(E)|^2 ,
\end{align}
up to possible channel-dependent normalization factors.
The CLD does not by itself determine $\Gamma(E)$, because
Eq.~\eqref{eq:cld_phase_relation} gives the phase of the transmission
amplitude rather than its modulus.
Nevertheless, the two quantities probe different aspects of the same
underlying scattering problem:
the greybody factor probes the probability for transmission through the
black-hole effective potential, whereas the CLD probes the rearrangement of
the continuum spectrum.

\revtwo{From this viewpoint, the CLD is complementary to, rather than a
replacement for, a direct greybody-factor calculation.  It organizes the
phase part of the scattering information into spectral contributions, whereas
the transmission probability additionally requires the modulus of the
transmission amplitude.  In the present paper we restrict the spectral
analysis to the selected-pole terms and the 
complementary remainder defined in Eq.~\eqref{eq:cld_remainder_omega}.}
If the transmission amplitude is written as
\begin{align}
    t(E)
    =
    |t(E)|
    \exp[i\delta(E)],
\end{align}
then the CSM decomposition gives, at least formally,
\begin{align}
    t(E)
    =
    |t(E)|
    \prod_\nu e^{i\delta_\nu(E)}
    \prod_{\nu'} e^{-i\delta_{0,\nu'}(E)}
    \times e^{i\delta_0}.
    \label{eq:t_phase_decomposition}
\end{align}
This expression should be understood as a decomposition of the phase part of
the transmission amplitude, not of the transmission probability itself.

\revtwo{A quantitative greybody factor requires $|t(E)|$, and a direct
real-frequency calculation also provides the natural independent benchmark
for the phase itself.  Neither calculation is performed here.  The claim of
the present CLD analysis is correspondingly narrower: it displays how the
phase-density trace is partitioned among identifiable resonance poles and the
remainder within the complex-scaled spectral representation.  Related work
based on isolated Riesz clusters of the complex-scaled black-hole resolvent
develops this connection further at the level of transmission, greybody
factors, and source--observer response near exceptional
points~\cite{Morikawa:2026fzn}.}

\section{Conclusion}\label{sec:conclusion}

\revtwo{The central question of this paper is whether the $\Lambda$-dependent
global spectral response of a Schwarzschild--dS black hole can be understood
from the motion of isolated QNM poles alone, or whether the complementary
spectrum is essential.  We addressed this question within the complex scaling
method by extracting QNM poles and the reference-subtracted continuum level
density from the same non-Hermitian spectral problem.}

We first formulated the Schwarzschild--dS perturbation problem in a
Schr\"odinger-type form and applied complex scaling in the tortoise
coordinate.  The outgoing-wave boundary-value problem is thereby converted
into a non-Hermitian spectral problem in which isolated resonance eigenvalues
are identified with QNMs and the continuum is rotated into the complex plane.

We applied the construction to scalar, electromagnetic, and gravitational
perturbations for several positive values of $\Lambda$.  For small $\Lambda$,
the extracted low-lying QNM frequencies connect continuously to the
Schwarzschild values.  As $\Lambda$ increases, both $\re\omega$ and
$|\im\omega|$ decrease for the modes studied here, consistently with the
approach toward the near-Nariai regime.  Agreement with previous QNM
calculations validates the identification of the isolated CSM eigenvalues.

\revtwo{The same complex-scaled spectra were then used to construct the CLD and
to decompose it into selected QNM-pole terms and a precisely defined remainder.
Across the dS$_4$ examples, the selected poles provide natural location and
width scales for the main spectral structures, but they do not in general
reproduce the total CLD by themselves.  In the broad small-$\Lambda$ regime the
lowest pole follows the total feature comparatively well, whereas at larger
$\Lambda$ the remainder can no longer be treated as a negligible correction
and is oppositely signed in the displayed decomposition.  For $l=3$, the first
overtone also becomes an appreciable component in the region of the sharp
feature.  The main conclusion is therefore that the QNM trajectory and the
global spectral response are related but not interchangeable: the relative
role of the selected poles and the remainder changes with $\Lambda$.}

\revtwo{We separately tested the finite-basis stability of the CLD in a
representative $s=2$, $l=3$, $\Lambda=0.08$ case.  The position and overall
shape of the main spectral feature are stable under the tested variations of
the scaling angle and Gaussian-basis parameters, while a visible residual
dependence remains in the pointwise amplitude.  The present numerical claims
are therefore deliberately qualitative at the amplitude level.  In
particular, we do not quote a converged cancellation ratio between pole and
remainder contributions.}

\revtwo{In ordinary one-dimensional scattering theory, the CLD is related to
the derivative of a global scattering phase.  This gives the traced quantity a
clear scattering interpretation and explains why the pole--remainder
separation may be viewed as a decomposition of the phase density.  The CLD is
not a spatially resolved Green function and does not determine a waveform or a
greybody factor.  Nor do we claim here an independent numerical reconstruction
of the real-frequency phase; a direct scattering calculation would provide
that benchmark and would also be required for the transmission modulus.}

\revtwo{Exploratory extensions are collected in
Appendix~\ref{sec:application}.  They illustrate how the same CSM construction
encounters multiple continuum thresholds in a string-inspired coupled-channel
problem and how it extends to higher-dimensional dS black holes.  These
examples are not used to support the main dS$_4$ CLD conclusions.}

\revtwo{The general CLD construction and a proof-of-principle pole-resolved
example were already presented in Appendix~B of Ref.~\cite{Ogawa:2026veu}.
The contribution of the present paper is the extension to the two-horizon
Schwarzschild--dS problem, the systematic $\Lambda$ dependence across three
perturbation sectors, and the explicit comparison between selected QNM poles
and the complementary reference-subtracted spectrum, together with a direct
finite-basis stability test of the CLD itself.}

\section*{Code availability}
A Julia package related to the tortoise-coordinate construction used in this
work, \texttt{AutoTortoise.jl}, is publicly available at
\url{https://github.com/o-morikawa/AutoTortoise.jl}.
The version used for this work corresponds to release \texttt{v0.1.4}.
It provides utilities for horizon-aware construction of tortoise maps and
their inverse maps for static spherically symmetric metric functions of
Laurent-polynomial type.

\section*{Acknowledgements}
This work was partially supported by Japan Society for the Promotion of Science (JSPS) Grant-in-Aid for Scientific Research Grant Numbers JP25K17402 (O.M.), and 25H01267 (S.O.).
O.M.\ acknowledges the RIKEN Special Postdoctoral Researcher Program
and RIKEN FY2025 Incentive Research Projects.

\appendix
\section{\revtwo{Relation to the reference-subtracted single-pole diagnostic}}
\label{sec:cld-alternative-diagnostic}

\revtwo{The main text uses the pole–remainder decomposition of Eqs.~\eqref{eq:cld_total_omega}--\eqref{eq:cld_remainder_omega}, in which the pure contributions of the explicitly selected QNM poles are separated from the complementary remainder. Here we consider a different diagnostic, obtained by retaining a single selected pole in the interacting resolvent while keeping the complete reference subtraction. This construction coincides with the convention used in Appendix B of Ref.~\cite{Ogawa:2026veu}.}

\revtwo{Reference subtraction is intrinsic to the total CLD: it removes the
density already present in the reference problem.  Once the full-Hamiltonian
trace is truncated to a single QNM pole, however, there is no canonical
one-to-one pairing between that pole and the complete reference spectrum.
Attaching the entire reference term to the selected-pole approximation
therefore defines a diagnostic of the approximated resolvent, not a unique
pure modal contribution.}

\revtwo{Let $E^\theta_\nu$ be the complex-scaled eigenvalues of the full
Hamiltonian and let $E^\theta_{0,\nu'}$ be those of the reference Hamiltonian.
The total quantity is}
\begin{align}
    \revone{\vsub{\Delta}{tot}^{\theta}(\omega)}
    &\revone{=
    -\frac{2\omega}{\pi}
    \im
    \left[
    \sum_\nu
    \frac{1}{\omega^2-E^\theta_\nu}
    -
    \sum_{\nu'}
    \frac{1}{\omega^2-E^\theta_{0,\nu'}}
    \right].}
    \label{eq:cld_total_omega_appendix}
\end{align}
\revtwo{For an explicitly selected QNM pole $E_n^\theta$, we define the reference-subtracted single-pole diagnostic by replacing the full-Hamiltonian sum in Eq.~\eqref{eq:cld_total_omega_appendix} by that pole while keeping the complete reference subtraction,}
\begin{align}
    \revone{\Delta_{n,\mathrm{ref}}^{\theta}(\omega)}
    &\revone{=
    -\frac{2\omega}{\pi}
    \im
    \left[
    \frac{1}{\omega^2-E^\theta_{n}}
    -
    \sum_{\nu'}
    \frac{1}{\omega^2-E^\theta_{0,\nu'}}
    \right].}
    \label{eq:cld_n_ref_omega}
\end{align}
\revtwo{Equation~\eqref{eq:cld_n_ref_omega} is the CLD obtained when the interacting resolvent is approximated by a single selected pole while the reference problem is left unchanged. 
It therefore provides a diagnostic of how closely that single-pole approximation reproduces the total reference-subtracted spectral structure. It must not be identified with the pure pole contribution $\Delta_n^\theta$ of Eq.~\eqref{eq:cld_pole_omega}.
When more than one QNM pole is displayed, the corresponding single-pole diagnostics are evaluated independently. 
In particular, $\Delta^\theta_{0,\mathrm{ref}}$ and $\Delta^\theta_{1,\mathrm{ref}}$ are not additive, since each contains the complete reference subtraction.}



\revtwo{For $l=3$, where both the $n=0$ and $n=1$ poles are identifiable at $\theta=43^\circ$, Eq.~\eqref{eq:cld_n_ref_omega} is evaluated separately for the two poles. 
The following figures compare these single-pole diagnostics with the total CLD for the same parameter sets as in the main text.}

\begin{figure}[ht]
    \pagehspace
    \includegraphics[width=1.25\columnwidth]{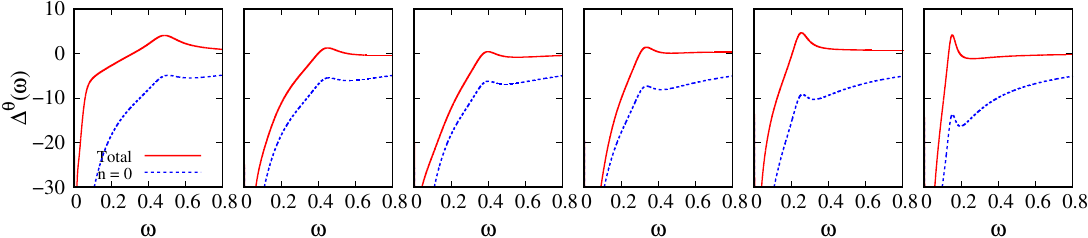}
    \caption{\revtwo{Alternative reference-subtracted single-pole diagnostic
    for $s=0$, $l=2$, and $\theta=43^\circ$.  The six panels correspond, from
    left to right, to~$\Lambda=0$, $0.02$, $0.04$, $0.06$, $0.08$, and $0.10$.
    The red solid curve denotes the total CLD of
    Eq.~\eqref{eq:cld_total_omega_appendix}.  The blue dashed curve denotes
    $\vsub{\Delta}{n=0,ref}^{\theta}$ of
    Eq.~\eqref{eq:cld_n_ref_omega}; it is a reference-subtracted single-pole
    approximation and not the pure $n=0$ pole contribution.}}
    \label{fig:cld_s0_l2_original}
\end{figure}

\begin{figure}[ht]
    \pagehspace
    \includegraphics[width=1.25\columnwidth]{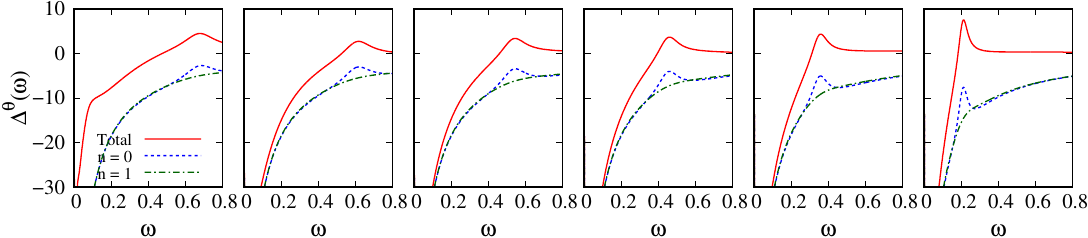}
    \caption{\revone{Same as Fig.~\ref{fig:cld_s0_l2_original}, but for
    $s=0$ and $l=3$.  The green dashed-dotted curve denotes the analogous
    $\vsub{\Delta}{n=1,ref}^{\theta}$ diagnostic.}}
    \label{fig:cld_s0_l3_original}
\end{figure}

\begin{figure}[ht]
    \pagehspace
    \includegraphics[width=1.25\columnwidth]{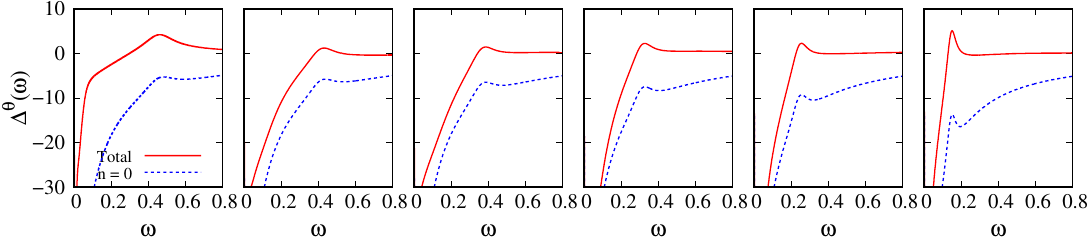}
    \caption{\revone{Same as Fig.~\ref{fig:cld_s0_l2_original}, but for
    $s=1$ and $l=2$.}}
    \label{fig:cld_s1_l2_original}
\end{figure}

\begin{figure}[ht]
    \pagehspace
    \includegraphics[width=1.25\columnwidth]{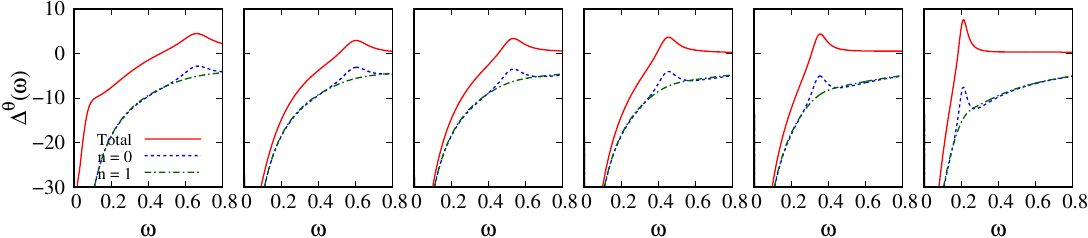}
    \caption{\revone{Same as Fig.~\ref{fig:cld_s0_l3_original}, but for
    $s=1$ and $l=3$.}}
    \label{fig:cld_s1_l3_original}
\end{figure}

\begin{figure}[ht]
    \pagehspace
    \includegraphics[width=1.25\columnwidth]{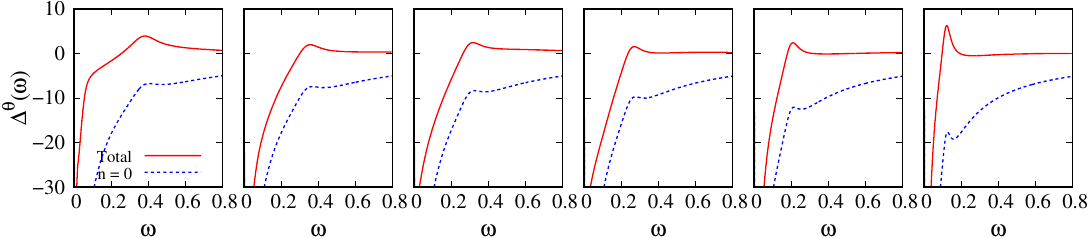}
    \caption{\revone{Same as Fig.~\ref{fig:cld_s0_l2_original}, but for
    $s=2$ and $l=2$.}}
    \label{fig:cld_s2_l2_original}
\end{figure}

\begin{figure}[ht]
    \pagehspace
    \includegraphics[width=1.25\columnwidth]{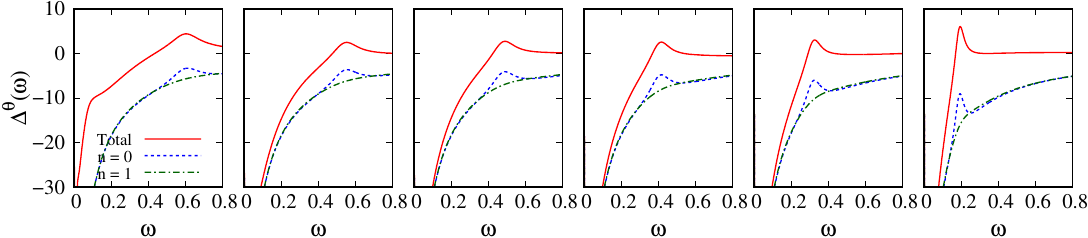}
    \caption{\revone{Same as Fig.~\ref{fig:cld_s0_l3_original}, but for
    $s=2$ and $l=3$.}}
    \label{fig:cld_s2_l3_original}
\end{figure}

\revone{For the representative scalar case with $s=0$ and $l=2$, the total
CLD has a broad structure at small $\Lambda$ and develops a sharper feature
at lower frequency as $\Lambda$ increases.  The alternative
$\vsub{\Delta}{n=0,ref}^{\theta}$ diagnostic follows much of this
qualitative evolution.  This is consistent with the movement of the lowest
QNM frequency in Table~\ref{tab:result_ds}: both $\re\omega_{n=0}$ and
$|\im\omega_{n=0}|$ decrease toward the near-Nariai regime.  Nevertheless,
the comparison here concerns a reference-subtracted single-pole approximation;
it does not by itself determine whether the pure pole term reproduces the
height, sign, or full width of the total CLD.  Those modal questions are
addressed by the main-text decomposition.}

\revone{The same distinction applies to the other spin and angular-momentum sectors.  
The Appendix figures show whether replacing the interacting trace by one selected pole retains the qualitative reference-subtracted structure, 
\revtwo{whereas Figs.~\ref{fig:cld_s0_l2}--\ref{fig:cld_s2_l3} separate the explicitly selected pure-pole contributions from the complementary remainder. 
The Appendix figures instead test, one pole at a time, how well the corresponding reference-subtracted single-pole approximation follows the total CLD.}
}

\revtwo{The main-text and Appendix constructions use the same total reference-subtracted CLD but answer different diagnostic questions:
the main-text decomposition isolates pure QNM-pole contributions, whereas the present Appendix compares the total CLD with reference-subtracted single-pole approximations evaluated separately for each selected pole.}

\section{\revtwo{Exploratory extensions of the framework}}\label{sec:application}
\revtwo{This appendix collects two exploratory extensions whose purpose is
to test how the same complex-scaling construction generalizes when the
spectral problem becomes more complicated.  No conclusion about the numerical
stability or physical interpretation of the main dS$_4$ CLD analysis relies on
these examples.}
\subsection{Coupled-channel systems: toward stringy black holes}
Recent work on the two-dimensional Mandal--Sengupta--Wadia (MSW) black hole shows that intrinsic
metric--dilaton perturbations can be reduced not to a single Schr\"odinger
equation but to a coupled matrix Schr\"odinger problem,
\begin{align}
    -\frac{\rmd^2}{\rmd r_*^2}\Phi(r_*)
    +
    \vsub{V}{eff}(r_*)\Phi(r_*)
    =
    \omega^2 \Phi(r_*),
    \label{eq:string_coupled_eq}
\end{align}
with a $2\times2$ effective potential matrix.
Here, $\Phi$ is a two-component field describing the coupled
metric--dilaton perturbation sector.
In the field basis used in Ref.~\cite{Bian:2026crb},\footnote{%
This basis is obtained by redefining the original metric and dilaton
perturbations so that Eq.~\eqref{eq:string_coupled_eq} has a
Schr\"odinger-type form without first-derivative terms.}
the elements of the effective potential are given by
\begin{align}
    \vsub{V}{eff}^{11,22}
    &= \frac{8\Lambda^2 e^{2\Lambda r_*}}{M\Lambda + e^{2\Lambda r_*}} ,&
    \vsub{V}{eff}^{12}
    &= \frac{4\vsub{V}{eff}^{11}}{8-\vsub{V}{eff}^{11}/\Lambda^2} ,&
    \vsub{V}{eff}^{21}
    &= \frac{1}{128} \vsub{V}{eff}^{11}
    \left(8-\frac{\vsub{V}{eff}^{11}}{\Lambda^2}\right)
    \left(32-\frac{\vsub{V}{eff}^{11}}{\Lambda^2}\right).
    \label{eq:string_pot}
\end{align}

Usually, the kinetic term and the overlap matrix can be evaluated
analytically for the basis functions used in the numerical variational method,
whereas the potential matrix elements require numerical integration.
Here, let us adopt the Polynomial $\times$ real range Gaussian basis.
Using Kummer's transformation for the confluent hypergeometric function, we
obtain
\begin{align}
    \int_{-\infty}^\infty \rmd r_*\, r_*^n e^{-\alpha r_*^2} \vsub{V}{eff}^{12}
   &= 
\begin{cases}
\frac{
4\sqrt{\pi}\,k!\,\Lambda
}{
M
}
\alpha^{-k-\frac{1}{2}}
\exp\!\left(\frac{\Lambda^2}{\alpha}\right)
L_k^{-\frac{1}{2}}
\!\left(
-\frac{\Lambda^2}{\alpha}
\right) & \text{for $n=2k$} ,
\\
\frac{
4\sqrt{\pi}\,k!\,\Lambda^2
}{
M
}
\alpha^{-k-\frac{3}{2}}
\exp\!\left(\frac{\Lambda^2}{\alpha}\right)
L_k^{\frac{1}{2}}
\!\left(
-\frac{\Lambda^2}{\alpha}
\right) & \text{for $n=2k+1$} .
\end{cases}
  \label{eq:kummer_transf}
\end{align}
Here, $L_k^a(z)$ denotes the generalized Laguerre polynomial.
Thus, the potential matrix element associated with $\vsub{V}{eff}^{12}$ can
be evaluated analytically, without performing the numerical integration.\footnote{%
In the coupled-channel problem, some off-diagonal components of the effective
potential can become exponentially large in the original field basis, while
the eigenvalues of the potential matrix remain finite.
This indicates that the difficulty is not a physical divergence of the
potential, but a numerical conditioning problem associated with the choice of
channel basis.
A suitable channel rescaling or matrix balancing is therefore required before
performing the complex-scaled diagonalization. See Appendix~\ref{sec:balance}.}

\begin{figure}[htbp]
    \centering
    \includegraphics[width=0.75\columnwidth]{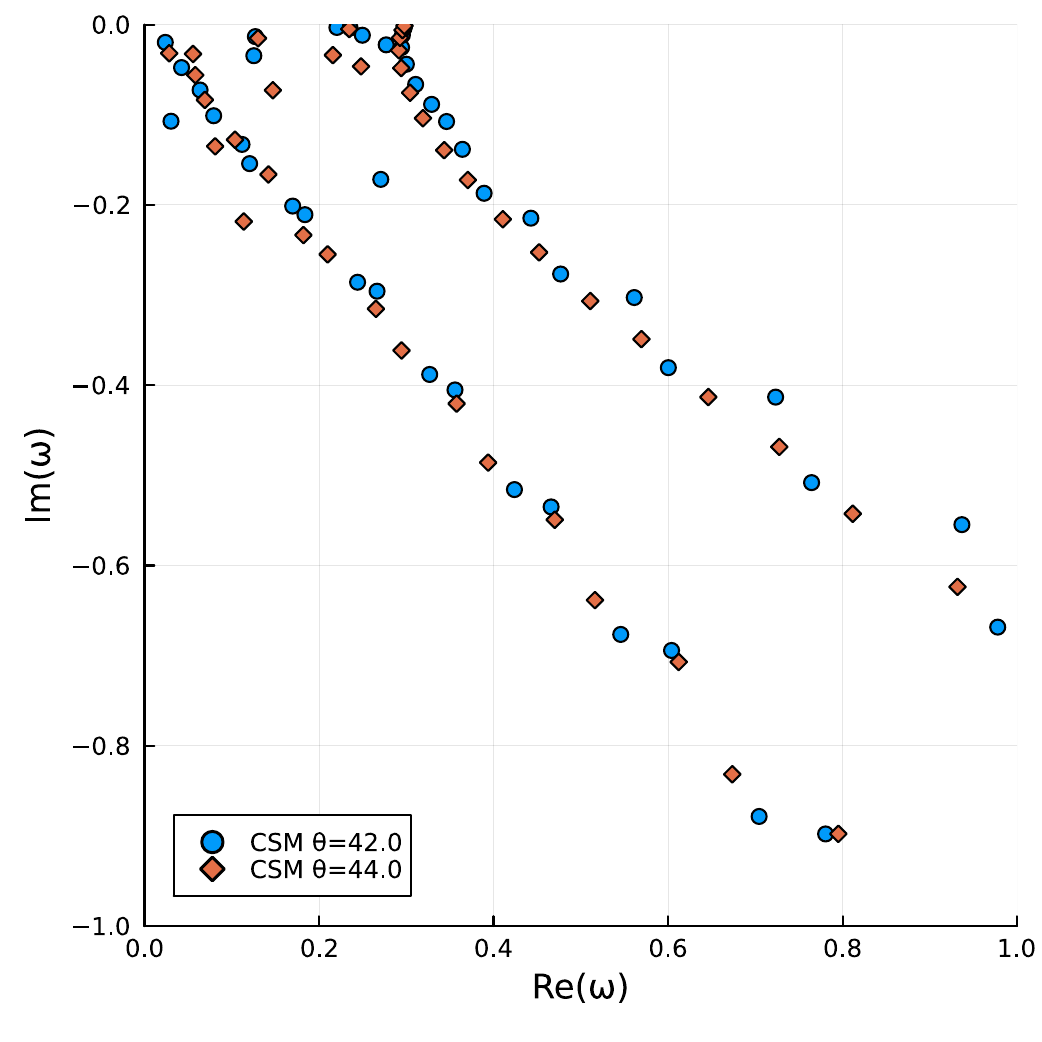}
    \caption{Complex-scaled spectrum for the coupled-channel MSW black-hole
    perturbation problem with $\Lambda=0.08$.
    \revone{No angular-momentum label is assigned to this two-dimensional
    system.}
    The scaling angle is $\theta=42^\circ$ and $44^\circ$.
    We set $i_{\max}=50$, $r_0=0.1$, and $r_{*\max}=80$.
    Because the asymptotic effective potential has two eigenvalue thresholds,
    two continuum branches are observed.
    Within the parameter sets examined in this work, no stable isolated
    resonance eigenvalue was identified for $\theta<45^\circ$.
    }
    \label{fig:stringy_coupled_spectrum}
\end{figure}

We next apply the CSM to the coupled-channel system
\eqref{eq:string_coupled_eq}.
A characteristic feature of this problem is that the asymptotic effective
potential has two eigenvalue thresholds.
Consequently, the complex-scaled spectrum contains two rotated continuum
branches, each starting from a different threshold.
This is in contrast to the single-channel Schwarzschild--dS examples
discussed above, where the continuum branch starts from a single threshold.
The result is shown in Fig.~\ref{fig:stringy_coupled_spectrum}.
The two lines show the corresponding complex-scaled continua
rotated by the angle $2\theta$.

To examine whether an isolated QNM eigenvalue can be exposed between the real
axis and the rotated continua, we performed a parameter scan over the basis
size, Gaussian range, integration range, and scaling angle.
The representative parameter sets used in this search are summarized in
Table~\ref{tab:stringy_parameter_sets}.
In all cases, we restricted the scaling angle to
\begin{align}
    \theta < \frac{\pi}{4},
\end{align}
which is the practical range of the present real-range Gaussian basis.

\begin{table}[htbp]
\centering
\caption{Representative parameter sets used in the CSM search for isolated resonance
eigenvalues in the coupled-channel MSW black-hole perturbation problem.
}
\label{tab:stringy_parameter_sets}
\begin{tabular}{ccccccc}
\toprule
Set & $i_{\max}$ & $p_{\max}$ & $r_{*0}$ & $r_{*\max}$ & $\theta$ & $\Lambda$ \\
\midrule
1 & 40 & 3 & 0.1 & 60 & $30^\circ$--$44^\circ$ & $0.01$--$0.1$ \\
2 & 40 & 3 & 0.1 & 80 & $30^\circ$--$44^\circ$ & $0.01$--$0.1$ \\
3 & 50 & 3 & 0.1 & 60 & $30^\circ$--$44^\circ$ & $0.01$--$0.1$ \\
4 & 50 & 3 & 0.1 & 80 & $30^\circ$--$44^\circ$ & $0.01$--$0.1$ \\
\bottomrule
\end{tabular}
\end{table}

Within this survey, we did not find a stable isolated eigenvalue that can be
identified as a QNM for $\theta<45^\circ$.
This should not be interpreted as evidence for the absence of QNMs in the
underlying stringy black-hole perturbation problem.
Rather, it indicates a limitation of the present real-range Gaussian CSM
implementation.
Indeed, the QNMs reported in Ref.~\cite{Bian:2026crb} are strongly damped in
the parameter region studied there.
For such modes, the corresponding position in the complex
$E=\omega^2$ plane can lie in a region that requires a larger rotation angle
to expose the pole from the continuum background.
However, the present basis representation becomes ill-conditioned as
$\theta$ approaches $\pi/4$, and therefore does not allow us to reliably
search for such poles beyond this angle.

The present result is nevertheless informative.
It shows that the coupled-channel CSM calculation correctly resolves the
multi-threshold continuum structure, while also revealing that the extraction
of highly damped QNM poles requires further stabilization.
Possible improvements include complex-range Gaussian bases, exterior complex
scaling, and channel-basis or matrix balancing of the type discussed in
Appendix~\ref{sec:balance}.
We therefore regard the stringy coupled-channel example as a useful
diagnostic problem for future extensions of the CSM framework, rather than as
a completed QNM computation in the present implementation.

\subsection{Higher-dimensional dS black holes}

As another natural direction, one may consider higher-dimensional
Schwarzschild--dS spacetimes.
In $d$ dimensions, the metric may be written as~\cite{Kodama:2003jz,Ishibashi:2003ap,Ishibashi:2011ws}\footnote{%
The black-hole mass parameter $M$ is
related to a horizon radius $r_0$ as
\begin{align}
    M = \frac{(d-2)A_{d-2}r_0^{d-3}}{16\pi G_d},
    \qquad
    A_{d-2} = \frac{2\pi^{(d-1)/2}}{\Gamma((d-1)/2)} ,
\end{align}
where $A_{d-2}$ denotes the area of the unit $(d-2)$-sphere.}
\begin{align}
    \rmd s^2 &= -f_d(r) \rmd t^2+f_d^{-1}(r)\rmd r^2+r^2 \rmd\Omega_{d-2},
    \label{eq:ds_d_metric}
    \\
    f_d(r) &= K-\lambda r^2-\frac{2M}{r^{d-3}},
    \label{eq:ds_d_metric_fn}
    \\
    \lambda &\equiv \frac{2\Lambda}{(d-1)(d-2)}.
\end{align}
$K$ denotes the sectional curvature of the $(d-2)$-dimensional base space.
In the present discussion, we mainly restrict ourselves to the case $K=1$.

For tensor-type perturbations on the $(d-2)$-dimensional angular sector with~$d\geq5$, the
effective potential is given by
\begin{align}
    V_T(r)
    &=
    \frac{f_d(r)}{r^2}
    \left[
        l(l+d-3)
        +\frac{(d-2)r f'_d(r)}{2}
        +\frac{(d-2)(d-4)f_d(r)}{4}
    \right].
    \label{eq:ds_d_pot_tensor}
\end{align}
For vector-type perturbations, one has
\begin{align}
    V_V(r)
    &=
    \frac{f_d(r)}{r^2}
    \left[
        l(l+d-3)
        +(d-4)
        +\frac{d(d-2)f_d(r)}{4}
        -\frac{(d-2)r f'_d(r)}{2}
    \right].
    \label{eq:ds_d_pot_vec}
\end{align}
The scalar-type potential is considerably more complicated.

These expressions suggest that higher-dimensional dS black holes provide
a natural class of backgrounds in which the present CSM framework may be
extended.
At least for the tensor- and vector-type sectors, the perturbation equations
remain of Schr\"odinger type, so that the basic spectral strategy should carry
over in a relatively direct manner.

As a simple illustration of the higher-dimensional extension, we also perform
a trial calculation for five-dimensional Schwarzschild--dS (Schwarzschild--dS$_5$) black holes.
Here we do not attempt a systematic survey of the higher-dimensional
parameter space.
Rather, our purpose is to check whether the same CSM prescription used in the
four-dimensional analysis can be applied directly to the tensor- and
vector-type master equations.

\begin{figure}[ht]
    \centering
    \includegraphics[width=0.45\columnwidth]{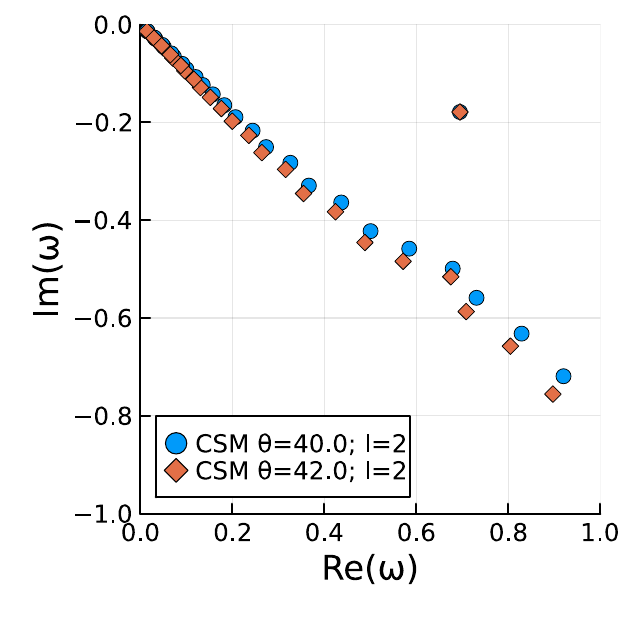}
    \includegraphics[width=0.45\columnwidth]{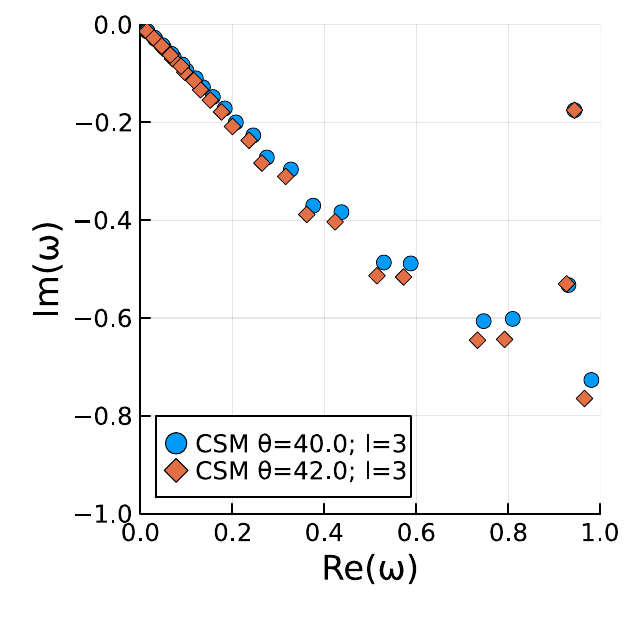}
    \caption{The Schwarzschild--dS$_5$ case for the tensor-type sector.
    $\Lambda=0.4$.
    The scaling angle is taken to be~$\theta=40^\circ$, $42^\circ$.
    We take  $i_{\max}=30$, $r_{*0}=0.1$, and $r_{*\max}=60$.
    $l=2$ on the left panel and $l=3$ on the right panel.}
    \label{fig:qnm_ds_d_s2}
\end{figure}

Figure~\ref{fig:qnm_ds_d_s2} shows the result for the Schwarzschild--dS$_5$
case in the tensor-type sector.
The complex-scaled spectra are shown for two angular momenta, $l=2$ and
$l=3$, and for two scaling angles, $\theta=40^\circ$ and $42^\circ$.
As in the four-dimensional case, the complex-scaled continuum is rotated in
the complex energy plane, and resonance candidates are identified as isolated
eigenvalues that are separated from the rotated continuum and remain stable
under a moderate change of the scaling angle.

\begin{figure}[ht]
    \centering
    \includegraphics[width=0.45\columnwidth]{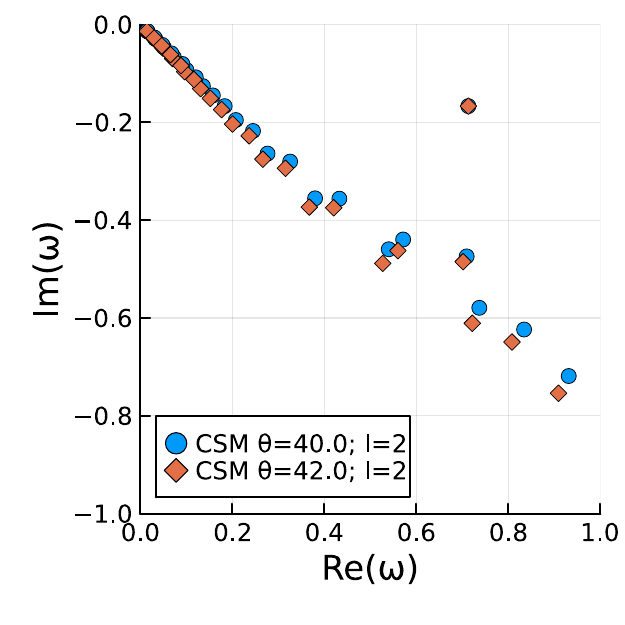}
    \includegraphics[width=0.45\columnwidth]{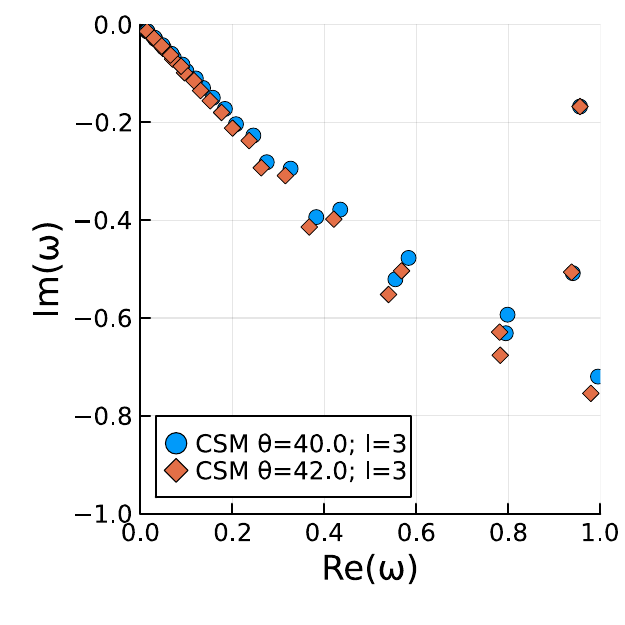}
    \caption{The Schwarzschild--dS$_5$ case for the vector-type sector.
    $\Lambda=0.4$.
    The scaling angle is taken to be~$\theta=40^\circ$, $42^\circ$.
    We take  $i_{\max}=30$, $r_{*0}=0.1$, and $r_{*\max}=60$.
    $l=2$ on the left panel and $l=3$ on the right panel.}
    \label{fig:qnm_ds_d_s1}
\end{figure}

Figure~\ref{fig:qnm_ds_d_s1} shows the corresponding example for the
Schwarzschild--dS$_5$ case in the vector-type sector.
Again, the same qualitative spectral structure is observed.
The appearance of stable isolated eigenvalues in these trial calculations
suggests that the CSM strategy can be extended to higher-dimensional
Schwarzschild--dS backgrounds at least for the tensor- and vector-type
sectors.
The present numerical results are in good agreement with those reported
by Konoplya~\cite{Konoplya:2003dd}.\footnote{%
When comparing the numerical values, we take into account the difference in
the normalization of the cosmological-constant parameter.  The cosmological
constant in the present convention is related to that used in
Ref.~\cite{Konoplya:2003dd} by~$\Lambda=2\vsub{\Lambda}{Konoplya}.$}
A detailed convergence analysis and a systematic comparison among different
dimensions, angular momenta, and perturbation sectors are left for future
work.

\section{Exponentially growing matrix blocks in diagonalization}\label{sec:balance}

Let us consider the generalized complex eigenvalue problem
\begin{align}
    H c = E N c ,
    \label{eq:generalized-evp-gaussian}
\end{align}
or, equivalently,
\begin{align}
    \sum_j
    \left[
        H_{ij}-E N_{ij}
    \right] c_j
    =0 .
\end{align}
For a two-channel problem, the Hamiltonian matrix has the block structure
\begin{align}
    H
    =
    \begin{pmatrix}
        T_{11}+V_{11} & V_{12} \\
        V_{21}        & T_{22}+V_{22}
    \end{pmatrix}.
\end{align}
In coupled-channel problems such as string-inspired black-hole perturbations,
some off-diagonal blocks can become exponentially large in the original field
basis.
For example, as seen in Eq.~\eqref{eq:kummer_transf}, the matrix element
associated with $V_{12}$ contains the factor
\begin{align}
    \exp\!\left(\frac{\Lambda^2}{\alpha}\right),
\end{align}
which can become very large for small Gaussian range parameters~$\alpha$.
This does not necessarily indicate a physical divergence of the effective
potential.
Rather, it can reflect a poor conditioning of the chosen channel basis.
In such a situation, the direct diagonalization of Eq.~\eqref{eq:generalized-evp-gaussian}
becomes numerically unstable.

\subsection{Block balancing}

A simple stabilization strategy is to apply a similarity transformation at the
matrix level.
We introduce the block scaling matrix
\begin{align}
    S
    =
    \begin{pmatrix}
        I & 0 \\
        0 & s I
    \end{pmatrix},
    \qquad
    s>0 ,
\end{align}
and write
\begin{align}
    c = S \tilde{c}.
\end{align}
Then Eq.~\eqref{eq:generalized-evp-gaussian} becomes
\begin{align}
    \left(
        S^{-1} H S
        -
        E S^{-1} N S
    \right)
    \tilde{c}
    =
    0 .
\end{align}
The scaled Hamiltonian is
\begin{align}
    S^{-1} H S
    =
    \begin{pmatrix}
        T_{11}+V_{11} & s V_{12} \\
        s^{-1}V_{21} & T_{22}+V_{22}
    \end{pmatrix}.
\end{align}
The eigenvalues are unchanged by this similarity transformation, while the
conditioning of the matrix problem can be improved.

A natural choice is to balance the norms of the two off-diagonal blocks:
\begin{align}
    s
    =
    \sqrt{
        \frac{
            \lVert V_{21}\rVert
        }{
            \lVert V_{12}\rVert
        }
    } .
\end{align}
Alternatively, if the exponential growth is dominated by the factor
$\exp(\Lambda^2/\alpha)$, one may use a scale estimate such as
\begin{align}
    s
    \sim
    \exp\!\left(
        -\frac{\Lambda^2}{\alpha_{\rm eff}}
    \right),
\end{align}
where $\alpha_{\rm eff}$ is a representative Gaussian range parameter.
The norm-balancing prescription is usually more robust, because it uses the
actual matrix elements after discretization.

\subsection{Channel-basis balancing}

A more structural approach is to change the channel basis before
discretization.
Consider the coupled Schr\"odinger-type equation
\begin{align}
    -
    \frac{\rmd^2}{\rmd r_*^2}
    \Phi(r_*)
    +
    V(r_*)\Phi(r_*)
    =
    E\Phi(r_*) .
\end{align}
We introduce a local channel rescaling
\begin{align}
    \Phi(r_*)
    =
    D(r_*)\Xi(r_*),
    \qquad
    D(r_*)
    =
    \begin{pmatrix}
        e^{\Lambda r_*} & 0 \\
        0               & e^{-\Lambda r_*}
    \end{pmatrix}.
\end{align}
Substituting this into the original equation and multiplying by $D^{-1}$ from
the left, we obtain
\begin{align}
    -
    \frac{\rmd^2}{\rmd r_*^2}
    \Xi
    -
    2D^{-1}D'
    \frac{\rmd}{\rmd r_*}
    \Xi
    +
    \left(
        D^{-1}VD
        -
        D^{-1}D''
    \right)
    \Xi
    =
    E\Xi .
    \label{eq:channel-balanced-eq}
\end{align}
Thus, the exponentially unbalanced off-diagonal potential is replaced by
\begin{align}
    V_{12}
    \mapsto
    e^{-2\Lambda r_*} V_{12},
    \qquad
    V_{21}
    \mapsto
    e^{2\Lambda r_*} V_{21}.
\end{align}
This can reduce the exponential hierarchy between the two off-diagonal
blocks.

The price of this transformation is the appearance of a first-derivative
coupling term,
\begin{align}
    -2D^{-1}D'
    \frac{\rmd}{\rmd r_*}\Xi .
\end{align}
Therefore, the resulting problem is no longer in the standard
Schr\"odinger form without first derivatives.
Nevertheless, this formulation may be numerically preferable if the original
off-diagonal blocks are so unbalanced that the direct diagonalization becomes
ill-conditioned.

In practice, the matrix-level block balancing is the simplest first test,
because it does not modify the differential equation.
If the instability persists, the channel-basis balancing of
Eq.~\eqref{eq:channel-balanced-eq} provides a more systematic way to remove
the exponential hierarchy at the level of the continuum problem.

\bibliographystyle{utphys}
\bibliography{ref,ref_ewkb,ref_qnm,ref_res,ref_scattering,ref_riesz}
\end{document}